\documentclass[authoryear,preprint,review,12pt]{elsarticle}
\usepackage[lmargin=1in, rmargin=1in, tmargin=1in, bmargin=1in, nohead]{geometry}

\usepackage{latexsym,graphics,amscd,pb-diagram}
\usepackage{amsmath,amssymb,mathrsfs,bm,graphicx,empheq,multirow,subcaption}
\usepackage{booktabs}

\usepackage{siunitx}
\usepackage{mathtools}
\usepackage{amsthm}
\usepackage{rotating}

\usepackage{epsfig}
\usepackage{epsf}
\usepackage{psfrag}
\usepackage{caption}

\usepackage{color}
\usepackage[all,cmtip]{xy}

\usepackage{changepage}
\usepackage{xassoccnt}
\usepackage{lastpage}
\usepackage{hyperref}
\usepackage{float}
\usepackage{appendix}
\usepackage[ruled]{algorithm2e}

\usepackage{lineno}

\usepackage{algorithm2e}

\NewTotalDocumentCounter{appendixchapters}

\newcommand{\myvec}[1]%
                         {\stackrel{\raisebox{-2pt}[0pt][0pt]{\small$\rightharpoonup$}}{#1}}

\NewTotalDocumentCounter{totalfigures}
\NewTotalDocumentCounter{totaltables}
\NewTotalDocumentCounter{appendixchapters}
\DeclareAssociatedCounters{figure}{totalfigures}
\DeclareAssociatedCounters{table}{totaltables}
\newcommand{\totalpages}{\zref@extractdefault{LastPage}{page}{0}}

\journal{NeuroImage}

\begin{document}

\begin{frontmatter}
\title{Quantification of Planar Cortical Magnification with Optimal Transport and Topological Smoothing}
\author[asu]{Yujian Xiong}
\author[asu]{Negar Jalili Mallak}
\author[asu]{Yanshuai Tu}
\author[nyu1,nyu2,nyu3]{Zhong-Lin Lu}
\author[asu]{\mbox{Yalin Wang}\corref{cor1}}
\address[asu] {School of Computing and Augmented Intelligence,\\
Arizona State University,
Tempe, AZ, USA}
\address[nyu1] {Division of Arts and Sciences, New York University Shanghai, Shanghai, China}
\address[nyu2] {Center for Neural Science and Department of Psychology, New York University, New York, NY, USA}
\address[nyu3] {NYU-ECNU Institute of Brain and Cognitive Science at NYU Shanghai, Shanghai, China}

\cortext[cor1]
{\emph{To be submitted to NeuroImage}
\vspace{10mm}
\begin{center}
\textbf{Abstract:}  203 words\\
\textbf{Title:}  96 characters\\
\textbf{Pages:}    \pageref{LastPage}  \\
\textbf{Figures:} \TotalValue{totalfigures} \\
\textbf{Tables:}  \TotalValue{totaltables} \\
\vspace{5mm}
Please address correspondence to: \\
\vspace{3mm}
Dr. Zhong-Lin Lu \\
Division of Arts and Sciences, \\
New York University Shanghai, Shanghai, China \\
Center for Neural Science and Department of Psychology, \\
New York University, New York, NY, USA \\
\textbf{E-mail:} zhonglin@nyu.edu \\
\vspace{3mm}
Dr. Yalin Wang \\
School of Computing and Augmented Intelligence\\
Arizona State University\\
P.O. Box 878809 \\
Tempe, AZ 85287, USA\\
\textbf{Phone:} (480) 965-6871  \\
\textbf{Fax:} (480) 965-2751  \\
\textbf{E-mail:} ylwang@asu.edu
\vspace{50mm}
\end{center}
}

\begin{abstract}
The human visual system exhibits non-uniform spatial resolution across the visual field, which is characterized by the cortical magnification factor (CMF) that reflects its anatomical basis. However, current approaches for quantifying CMF using retinotopic maps derived from BOLD functional magnetic resonance imaging (fMRI) are limited by the inherent low signal-to-noise ratio of fMRI data and inaccuracies in the topological relationships of the retinotopic maps. In this study, we introduced a new pipeline to quantify planar CMF from retinotopic maps generated from the population receptive field (pRF) model. The pipeline projected the 3D pRF solutions onto a 2D planar disk, using optimal transport (OT) to preserve local cortical surface areas, and applied topological smoothing to ensure that the resulting retinotopic maps maintain their topology. We then estimated 2D CMF maps from the projected retinotopic maps on the planar disk using the 1-ring patch method. Applying this pipeline to the Human Connectome Project (HCP) 7T dataset, we revealed previously unobserved CMF patterns across the visual field and demonstrated individual differences among the 181 subjects. The pipeline was further validated on the New York University (NYU) 3T dataset, showing reliable and repeatable results. Our study provided new analytical methods and offered novel insights into visual processing.
\end{abstract}

\end{frontmatter}

\newpage
\section{Introduction}\label{sec:1}

\begin{figure*}[t]
\begin{center}
\begin{tabular}{c}
\includegraphics[width=0.95\linewidth]{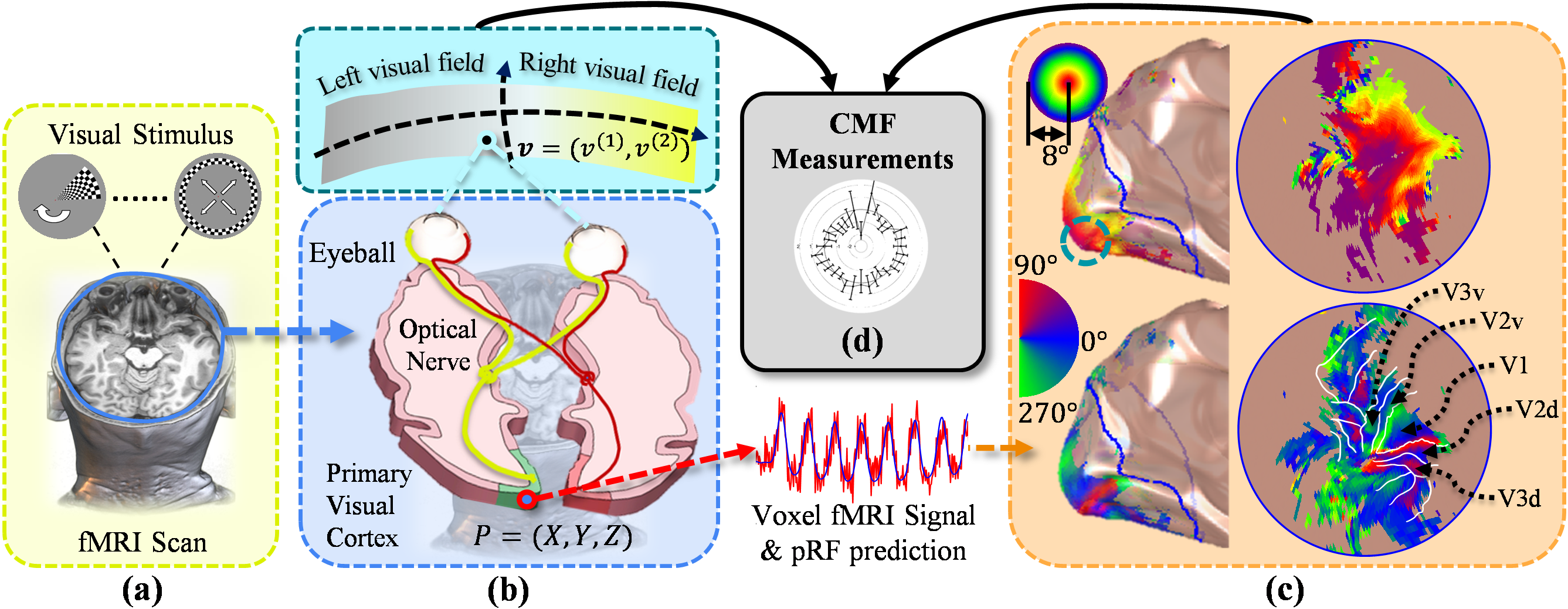}
\end{tabular}
\end{center}
\caption 
{ \label{fig:overview_prf}
\textbf{Standard retinotopic mapping through fMRI, pRF modeling and CMF measurements: } (a) Typical experimental setting with pRF visual stimuli including rotating wedges, expanding rings, and moving bars with checkerboard patterns. (b) These stimuli systematically activate the visual field and its cortical representation in the primary visual cortex. (c) vertex-level BOLD fMRI signals are modeled with the pRF method to estimate receptive field locations; the results are projected onto the cortical surface mesh, displaying eccentricity and polar angle for each surface vertex with labeled anatomical references for vertical and horizontal meridians. (d) Using the surface-based pRF parameters, we could measure 1D CMF (on certain vertical or horizontal directions) or 2D CMF (see Fig.~\ref{fig:general}).} 
\end{figure*}

The mammalian visual cortex contains multiple representations of distinct regions of the visual field~\citep{horton1991representation, cowey1964projection}. Research on retinotopic representation, based on the activity and anatomy of the visual cortex, has generated considerable interest~\citep{sereno1995borders, Engel1997, Swindale2000, warnking2002fmri, wandell2007visual, dougherty2003visual}. In numerous investigations, retinotopic mapping of the human brain has relied on blood oxygen level-dependent (BOLD) functional magnetic resonance imaging (fMRI)~\citep{schneider1993functional, sereno1995borders, tootell1995functional, DeYoe19031996, Engel1994, Engel1997, brewer2005visual, dumoulin2008population, tootell1998functional} and employed the population receptive field (pRF) model~\citep{kay2013compressive,Kay2018,waz2024qprf} to estimate the centers and sizes of the receptive fields for each vertex on the visual cortical surface, as shown in Fig.~\ref{fig:overview_prf}. The results of these analyses have significantly improved our understanding of the human visual system~\citep{Schwarzkopf2011, Wandell2011, brewer2014visual, Michel2013, Morland2001, Kammen:NIMG16, dumoulin2008population, benson2018bayesian, lerma2020validation, lage2020investigating}. 

The cortical magnification factor (CMF), a measure of the amount of cortical tissue in the visual cortex that is devoted to processing a given degree of visual field, has been shown to exhibit unique developmental trajectories that are critical to understanding neurological and psychological functionalities~\citep{hazlett2011early, ecker2013brain, mensen2017development, palaniyappan2011regional, cheng2021genetic, bois2015cortical, rimol2012cortical, vuoksimaa2015genetic, zacharopoulos2020cortical, vuoksimaa2016bigger}. Vision scientists have used CMF to investigate many aspects of processing resources in the cortex, including neural connection density, synaptic efficiency, and brain plasticity.

It has been widely used as a metric to quantify the relationship between visual acuity and cortical anatomy~\citep{cowey1974human, lage2020investigating, bordier2015quantitative, qiu2006estimating, duncan2003cortical, song2015neural,zeidman2018bayesian,yu2017intrinsic}, offering promising insights into perceptual asymmetries such as horizontal-vertical asymmetry (HVA)~\citep{benson2021cortical, himmelberg2021cross, himmelberg2023polar, himmelberg2023comparing}, differences between specific visual cortical areas~\citep{cowey1974human, harvey2011relationship, silva2018radial, kupers2022asymmetries} and individual differences~\citep{himmelberg2022linking, kupers2022asymmetries, song2022linking, benson2021cortical, himmelberg2021cross}.

Despite its widespread adoption~\citep{virsu1996cortical, horton1991representation, duncan2003cortical, song2015neural, hussain2015estimation, bordier2015quantitative, lage2020investigating, benson2021cortical, himmelberg2021cross, himmelberg2023polar, himmelberg2023comparing}, traditional approaches for estimating CMF faced notable limitations in both precision and scalability. Broadly, two primary approaches have been employed. The first was a model-based method, in which a 2D angle-preserving algebraic model or its enhanced non-conformal variant was fitted to the retinotopic maps. CMF was then derived as a function of eccentricity and polar angle from the best-fitting model~\citep{Daniel1961, Schwartz1977, Schwartz1980, schira2007two, schira2009foveal}. However, studies have found that human retinotopic maps were quasiconformal~\citep{Ta2022}. The log-transformation function~\citep{Schwartz1977, Schwartz1980}, along with its enhanced version incorporating a double-sech shearing function~\citep{schira2007two}, can only approximate the average shape of the retinotopic maps. These models cannot account for asymmetries and local variations among observers, and therefore, they cannot be used to accurately reconstruct the original retinotopic maps. As a result, they can lead to significant inaccuracies in CMF estimates.

As an alternative to model-based methods, data-driven empirical approaches have gained popularity due to the limitations in precision and generalizability of the former. In the empirical estimation of CMF, a short line was delineated on the cortical surface, and the corresponding segment was identified in the visual field. The CMF was then calculated as the ratio between the cortical distance and the corresponding distance in visual space~\citep{cowey1974human}. The process typically involved two key steps: (1) extracting the population receptive field for each vertex on the cortical surface, including its center and size, based on vertex-level fMRI signals using the pRF model and numerical estimation techniques~\citep{dumoulin2008population, lage2020investigating, lerma2020validation}; and (2) mapping these cortical locations to their respective receptive field centers in the visual field~\citep{harvey2011relationship}. In this process, the cortical distance $d$ (in millimeters) between neighboring locations was measured along the cortical surface,  while the corresponding change in visual field location, $\Delta \rho$ (degrees of visual angle), was derived from pRF estimates. The CMF was then calculated as the ratio $c = \frac{d}{\Delta \rho}$~\citep{harvey2011relationship}. 

However, empirical CMF estimation faced two key challenges. \textbf{(1). One Dimensional CMF Measures.} Existing methods typically computed retinotopic maps on a per-vertex basis, resulting in one-dimensional (1D) CMF measures that did not fully capture the inherently two-dimensional (2D) nature of retinotopy. These 1D measurements, often averaged or computed along a single direction, can only reflect asymmetries by comparing the CMF values in different orientations~\citep{kupers2022asymmetries, silva2018radial}. Such an approach lacked detailed spatially specific information, particularly near the fovea, where high-resolution mapping was most critical. A two-dimensional CMF representation was necessary to overcome this limitation. By providing a spatially continuous mapping of the visual field onto the cortex, 2D CMF enabled a more nuanced understanding of cortical magnification, especially in high-acuity regions with densely packed receptive fields. However, realizing accurate 2D CMF estimation introduced a second major challenge. \textbf{(2). Topological Violations.} Topological violations occurred when mapping from visual space to cortical surface failed to preserve local neighborhood relationships, particularly in regions with complex geometry, such as the fovea. The low signal-to-noise ratio and limited spatial resolution in fMRI often led to such violations in pRF estimates, distorting the smooth and continuous representations of the visual space on the cortex~\citep{Tu:Plos2021}. Consequently, CMF measurements may become distorted or unreliable~\citep{dumoulin2008population, benson2018human, Tu:Plos2021}. To mitigate these issues, several optimization methods have been proposed, including reliability-based analyses of the size of the perceptual field~\citep{lage2020investigating, lerma2020validation}, model-free back projections to recover perceptual shapes more accurately~\citep{merkel2018spatial, merkel2020modulating, lerma2021population}, and various smoothing techniques such as Gaussian~\citep{Warnking2002, bordier2015quantitative}, Laplacian~\citep{zeidman2018bayesian}, and mesh-spectrum-based smoothing~\citep{qiu2006estimating}. While these techniques improved the robustness of pRF-derived maps, none directly resolve topological violations --- except for our prior work~\citep{Ta2014, Tu:Plos2021, Ta2022}. Such violations remained a significant barrier to accurate CMF measurements, particularly on highly curved cortical surfaces~\citep{Ta2022, Tu:Plos2021}.

\begin{figure*}[t]
\begin{center}
\begin{tabular}{c}
\includegraphics[width=0.8\linewidth]{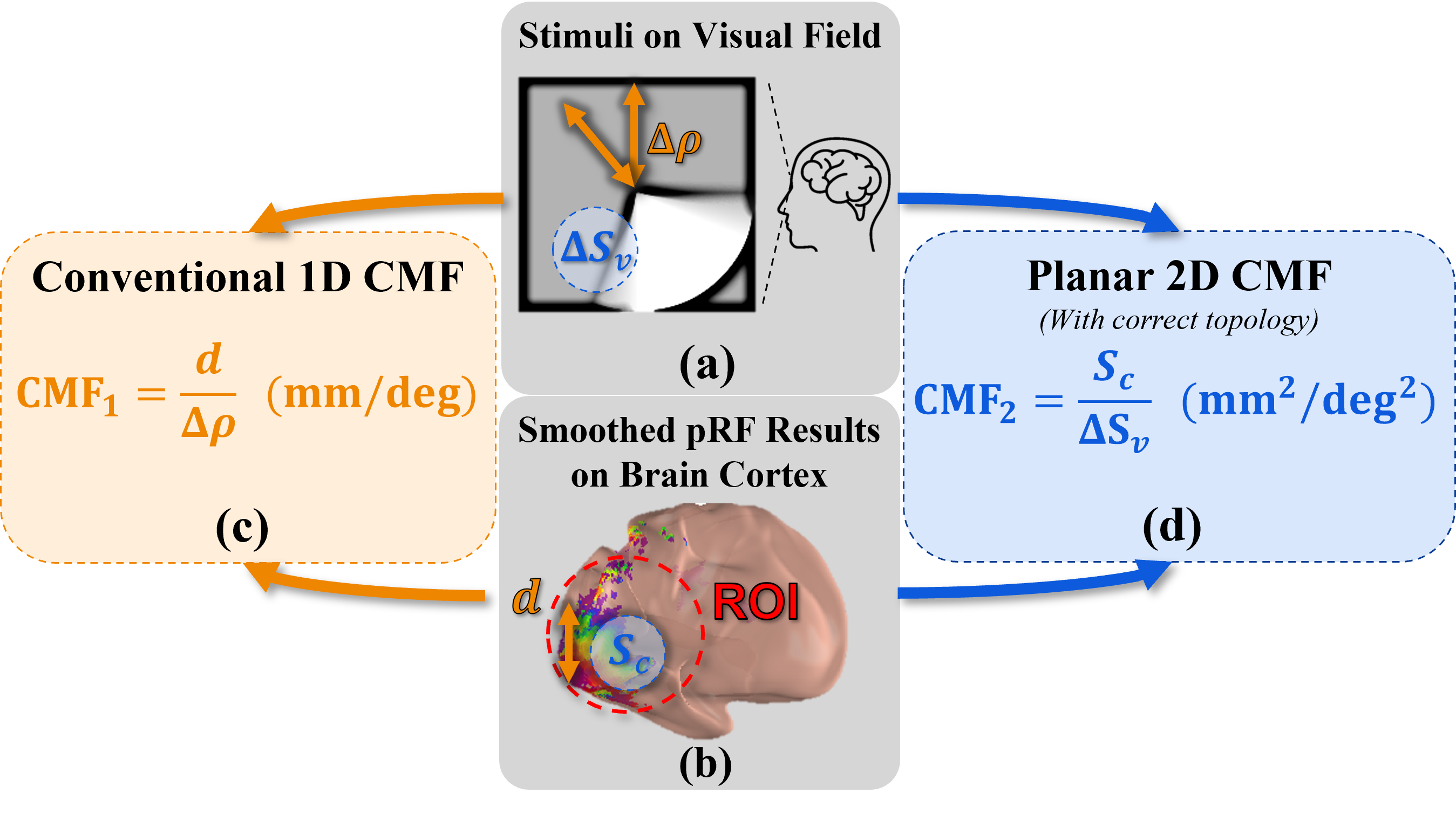}
\end{tabular}
\end{center}
\vspace*{-7mm}
\caption 
{ \label{fig:general}
\textbf{Comparison of CMF measurements: } (a) The visual field of pRF stimulus viewed by the subjects. (b) The decoded pRF parameters for each vertex on brain cortex, which gives the corresponding receptive center on visual field for vertices. (c) Conventional CMF measurements that approximate CMF by the division between the cortical vertex distance $d$ and visual center distance $\Delta \rho$ on certain vertical or horizontal direction. (d) Our proposed planar CMF measurements that directly calculate the CMF for each vertex through the division between the cortical area $S_c$ of the 1-ring patch around certain vertex and corresponding visual field area $\Delta S_v$. This is only possible when we obtain a topological correct pRF results.}
\end{figure*}

To address the challenges in CMF estimation, we proposed a novel pipeline that applied optimal transport (OT)~\citep{bonnotte2013knothe} and topology-preserving smoothing~\citep{Tu:Plos2021} to the retinotopic maps derived from the pRF model. As a mathematical optimization technique, OT identified the most efficient way to ``transport" quantities between specific locations on two surfaces and can be used to generate area-preserving surface mapping. Expanding upon our previous research efforts~\citep{Su:TPAMI2015,Shi2019}, where we successfully relocated individual vertices on a 3D surface onto a 2D planar disk, we extended this approach to obtain an area-preserving parameterized disk from the cortical surface. With these techniques, our topology-preserving smoothing algorithm ensured that the geometric relationships in the retinotopic maps remained intact~\citep{Tu:Plos2021, tu2022protocol}. With a topological correct quantification of retinotopic maps, we can directly compute the CMF using the 1-ring patch method: for each vertex on the cortical surface, a neighboring patch is defined and its corresponding ``dual patch" was located in the visual field. The areas of these dual patches were then used to compute the local 2D CMF value at each vertex. The key differences between conventional 1D and the proposed 2D CMF measurements are shown in Fig.~\ref{fig:general}

Our pipeline preserved topological structure while enabling detailed quantification of cortical magnification across the visual field. Unlike previous conformal mapping approaches, our OT-based method improved area correspondence between the cortical surface and the mapped disk, while still preserving both topological and angle relationships. This enhancement allowed for accurate, region-specific CMF estimation based on area measurements, providing valuable insights into the relationship between cortical magnification and visual behavior. We validated the effectiveness of our method using real retinotopic data from the Human Connectome Project (HCP)~\citep{benson2018human} and New York University (NYU)~\citep{himmelberg2021cross}. Our results revealed previously undetected CMF patterns across the visual field and highlighted individual differences among subjects.

\section{Methods and Materials}\label{sec:Methods}

\subsection{Human Retinotopic Data}

In this study, we used two published human retinotopic datasets.

\textit{HCP Dataset.} The publicly available retinotopic data from the Human Connectome Project (HCP) can be accessed from (\href{https://www.humanconnectome.org/study/hcp-young-adult}{www.humanconnectome.org/study/hcp-young-adult}) \citep{uugurbil2013pushing, van2013wu} and (\href{https://osf.io/bw9ec/}{https://osf.io/bw9ec/})~\citep{benson2018human} or (\href{https://balsa.wustl.edu/study/show/9Zkk}{https://balsa.wustl.edu/study/show/9Zkk})\citep{van2017brain}. The current study uses a retinotopic dataset from 181 participants, each of whom underwent retinotopic mapping experiments using 7T fMRI scanning. The dataset includes the fitted pRF results for retinotopy, along with individual cortical surface atlases for each subject.

\textit{NYU Dataset.} The NYU Retinotopy Dataset~\citep{himmelberg2021cross} provides fMRI-based measurements of retinotopic maps in the human visual cortex, including data from 44 participants. Despite differences in protocols compared to the HCP dataset, the NYU data reliably reproduces key retinotopic properties across visual areas V1, V2, V3, and hV4. It confirms polar angle asymmetries in cortical magnification and demonstrates strong correlations in population receptive field (pRF) estimates, validating the robustness of fMRI-derived retinotopic maps. This dataset serves as a valuable benchmark for studying visual area organization and comparing to the HCP 7T Retinotopy Dataset.

\subsection{Retinotopic Mapping}
In this section, we introduce the background and key notations used in the retinotopic mapping experiment. Let the visual stimulus at position $v=(v^{\left(1\right)},v^{\left(2\right)})$ be denoted by $s\left(t,v\right)$, where $v^{\left(1\right)}$ represents the radial distance from the fovea (i.e., eccentricity) and $v^{\left(2\right)}$ represents the polar angle in the visual field. In each retinotopic region of the visual system, the visual stimulus activates a population of neurons.


The central task of retinotopic mapping is to determine the center $v$ and size $\sigma\in \mathbb{R}^+$ of the perceptual receptive field for each vertex $P=\left(X,Y,Z\right)\in\mathbb{R}^{3}$ on the cortical surface. BOLD fMRI provides a noninvasive way to obtain the signals needed to estimate $v$ and $\sigma$ for each point $P$, following this procedure: 

\begin{itemize}
\item \textbf{Stimulus Design:} Design the visual stimulus $s\left(t;v\right)$ such that it is uniquely defined by its location, i.e., $s\left(t;v_1\right)\neq s\left(t;v_2\right),\forall\ v_1\neq v_2$; 
\item \textbf{Stimulus Presentation and Data Acquisition:} Present the visual stimuli to a subject and record BOLD fMRI activation $y\left(t;P\right)$ at each vertex $P$ on the cortical surface.
\item \textbf{Population Receptive Field Estimation:} Determine the parameters of the population receptive field model, including the center location $v$ and the size of the receptive field $\sigma$, which is most likely to generate the fMRI signal $y(t;P)$ at each cortical vertex $P$. 
\end{itemize}


\begin{table}[t]
	\centering
	\begin{tabular}{|c|l|l|} \hline
		Region   Name & Left Hemisphere & Right Hemisphere    \\ \hline
		V1v &  $v^{(2)}=\theta+\pi$ & $v^{(2)}=\theta$   \\ \hline
		V1d & $v^{(2)}=\theta-\pi$ & $v^{(2)}=\theta$    \\ \hline
		V2v &$v^{(2)}=-\theta+2\pi$ & $v^{(2)}=-\theta+\pi$   \\ \hline
		V2d & $v^{(2)}=-\theta-2\pi$ &$v^{(2)}=-\theta-\pi$  \\ \hline
		V3v &$v^{(2)}=\theta+2\pi$ & $v^{(2)}=\theta+\pi$   \\ \hline
		V3d & $v^{(2)}=\theta-2\pi$ & $v^{(2)}=\theta-\pi$   \\
		 \hline
	
	\end{tabular}
	\caption{Polar angle transformations used to eliminate phase-jumping and correct changes in visual field signs in multiple visual areas~\citep{Tu:Plos2021}}\label{tab:polarangletransformation}
\end{table}

The polar angle $\theta$ in the visual field's polar coordinate system is discontinuous near the $x$ axis. However, the retinotopic maps are continuous. To remove these phase discontinuities and allow for a unified treatment of the retinotopic maps of the V1-V3 computation, we applied the polar angle transformations developed in our prior work~\citep{Tu:Plos2021}. These transformations (summarized in Table~\ref{tab:polarangletransformation}) generated extended polar angles, thereby providing a continuous mapping of retinotopic information across V1, V2, and V3.

\subsection{The Pipeline}

\begin{figure*}[t]
\begin{center}
\begin{tabular}{c}
\includegraphics[width=0.85\linewidth]{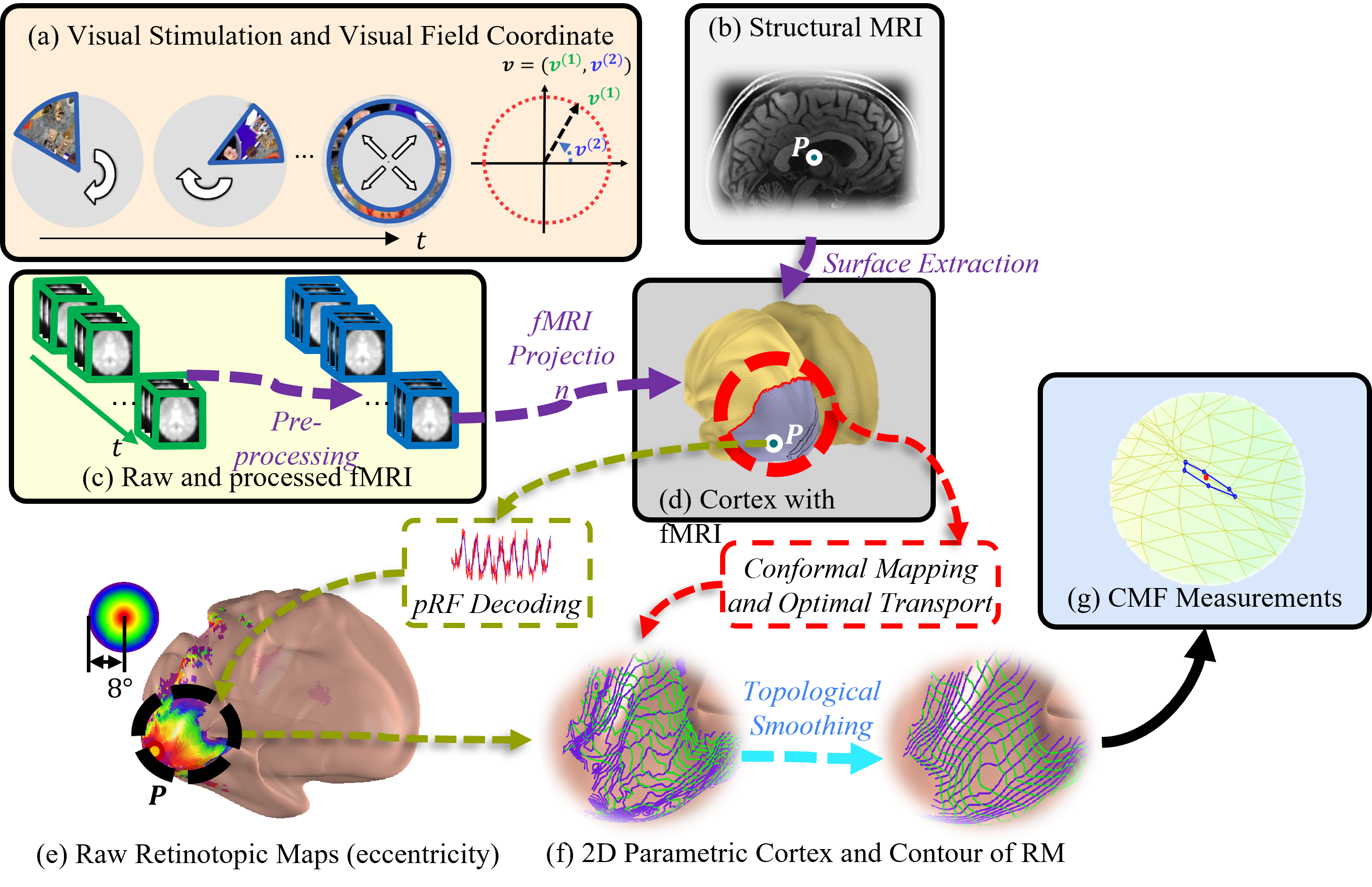}
\end{tabular}
\end{center}
\vspace*{-7mm}
\caption 
{ \label{fig:pipeline}
\textbf{Overview of pipeline: } Our proposed pipeline began with two primary data inputs: (a) visual stimuli and its coordinate system for the visual field, and (b) structural (anatomical) MR images, along with (c) BOLD fMRI scans. These data underwent preprocessing (c) to prepare them for analysis. Next, we reconstructed the cortical surface and projected the fMRI activation onto this surface (d). Using the population receptive field (pRF) model, we calculated retinotopic maps, extracting the receptive field center and size ($\sigma$) for each cortical vertex (e). We then applied the optimal transport to project the cortical surface onto a 2D planar disk, allowing us to visualize clear retinotopic maps and their contours (f). Finally, topology-preserving smoothing was applied to refine the retinotopic maps (f), enabling accurate CMF measurements (g).} 
\end{figure*}

\subsubsection{Overview}
The pipeline is illustrated in Fig. \ref{fig:pipeline}. It began with fMRI pre-processing and projections onto the cortical surface, followed by the application of area-preserving optimal transport (OT) to project the cortical surface onto a flattened 2D surface. Simultaneously, we computed raw retinotopic maps through pRF decoding and apply topology-preserving smoothing to correct any topological violations in the retinotopic maps. The smoothed retinotopic maps were then used for planar CMF computations with the 1-ring patch method. Finally, we merged the CMF maps from the two hemispheres to visualize CMF properties and assessed individual differences. This pipeline was applied to both the HCP and NYU datasets.

\subsubsection{fMRI Pre-Processing}
The primary goal of fMRI pre-processing was to accurately capture the brain's time series of activation in response to visual stimuli, ensuring that the process is both robust and sensitive. To achieve this, we employed the standardized HCP preprocessing pipeline. First, the cortical surface was extracted from structural MRI images using Freesurfer software and resampled. Then, the raw fMRI data from each imaging session were co-registered over time to correct for head movements and other motion artifacts. Finally, the co-registered fMRI data were projected onto the cortical surface, with spatial smoothing applied along the cortical surface to enhance signal quality. This process resulted in a resampled cortical surface and fMRI activation time series for each point on the cortical surface.

\subsubsection{pRF Decoding}
The population receptive field (pRF) model~\citep{dumoulin2008population,kay2013compressive} was used to determine the receptive field of each vertex $P = (X, Y, Z) \in \mathbb{R}^3 $ on the visual cortical surface, including its center location $v$ and size $\sigma$ in the visual field. For readers familiar with the pRF model, the visual coordinates in this study are related to the Cartesian coordinates via the transformation:
$x = v^{(1)} \cos{v^{(2)}}, y = v^{(1)} \sin{v^{(2)}} $. 

The pRF model predicts the fMRI signal for each vertex based on the following equation:
\begin{equation}
    \hat{y}(v,\sigma,n) = \beta(\int r(v^\prime;v,n)s(t,v^\prime)dv^\prime)^{n}*h(t)
\end{equation}

where $\beta$ is the activation level, $n$ is the exponent of the power function, $v'$ is a location in the visual field and $v$ is the center of the receptive field, $r(v^\prime;v,n)$ is the population receptive field model follows a Gaussian kernel: $r(v^\prime;v,n)=\exp{(-\frac{(x^\prime-x)^2 + (y^\prime-y)^2}{2\sigma^2})}$, and $h(t)$ is the hemodynamic function.

The center of the receptive field $v$ and the size of the population receptive field $\sigma$ are estimated by minimizing the least square difference between the predicted fMRI signal $\hat{y}(v,\sigma)$ and the measured signal $y(P)$, as shown in the following equation:

\begin{equation}
(v,\sigma,n)=\arg \min_{(v,\sigma,n)} |\hat{y}(v,\sigma)-y(P)|^2   
\end{equation}

This procedure is iterated across all vertices on the visual cortical surface to generate the raw retinotopic map. The goodness of fit for each vertex is assessed by calculating the $R^2$ value:

\begin{equation}
R^2=100(1-\int(\hat{y}-y)^2/\int y^2dt).
\end{equation}

\subsubsection{Optimal Transport}
Optimal transport (OT) is a mathematical framework used to determine the optimal method to transport the mass from one distribution to another, minimizing the associated transportation costs. In this approach, we considered two probability measures, $X(x, \mu)$ and $Y(y, v)$, which belong to the space of all Borel probability measures on a metric space $M$, denoted by $P(M)$. A measure-preserving mapping $\pi: X(x, \mu) \rightarrow Y(y, v)$ was considered optimal if it minimized the total transportation cost, which was determined by a given cost function $c: M \times M \rightarrow \mathbb{R}_0^+$. The optimal transport plan was represented by the mapping $\pi$.

The Wasserstein distance, also known as the Earth Mover's distance, is a metric used to quantify the similarity between two probability distributions. It is defined as the minimum cost of transporting the mass between the two distributions, with the cost being computed as the distance increased to the power of $p: c(x, \pi(x)) = d(x, \pi(x))^p$, where $d(x, \pi(x))$ is the distance between the point $x$ and its corresponding point $\pi(x)$ under the optimal transport plan $\pi$. The 2-D Wasserstein distance is obtained by mapping the probability measures $X(x, \mu)$ and $Y(y, v)$ to a unit hypersphere using volumetric harmonic maps, $\phi_1: M \rightarrow S^n$ and $\phi_2: N \rightarrow S^n$, where $S^n$ represents the canonical space parameterized by Cartesian coordinates. The empirical measures $X^\prime(x, \mu)$ and $Y^\prime(y, v)$ are then defined on the transformed supports $\phi_1(\Omega_X)$ and $\phi_2(\Omega_Y)$, respectively. The optimal transport map $T: X^\prime \rightarrow Y^\prime$ is computed to induce the 2-D Wasserstein distance $W_2(X^\prime, Y^\prime)$. The Wasserstein distance is then expressed as a sum over a relatively sparse point set $P(p, v)$, obtained by discretizing $Y^\prime$, and the distance between two points is measured as $d^2(m_i, p_j) = ||m_i - p_j||^2$, where $p_j$ represents the optimal mapping of the point $m_i$ under the optimal transport plan $\pi$~\citep{tu2020computing}. 

In the context of our pipeline, optimal transport can be used to establish the optimal area-preserving mapping between the original cortical surface and the parameterized planar surface, where each 2D point corresponds to a specific vertex on the cortical surface. Rather than simply using a conventional parametric planar surface derived from conformal mapping, we applied optimal transport to calculate the parametric coordinates of each vertex on the cortical surface to preserve the area and topology of the maps. The output was a modified parametric surface that aligns the retinotopic maps more accurately with much less area distortion, enabling CMF computation based on the division of areas.





    
    
    



\subsubsection{Optimal Transport}

Optimal transport (OT) provides a mathematical framework for determining the most efficient way to move mass from one distribution to another while minimizing the total transportation cost. In this study, two probability measures, $X(x, \mu)$ and $Y(y, v)$, are defined on a metric space $M$ belonging to the set of Borel probability measures $P(M)$. A measure-preserving mapping $\pi: X(x, \mu) \rightarrow Y(y, v)$ is considered optimal if it minimizes a transportation cost function $c: M \times M \rightarrow \mathbb{R}_0^+$, such that the total cost $\int c(x,\pi(x)) \, d\mu(x)$ is minimized. The resulting mapping $\pi$ is the optimal transport plan.

The Wasserstein distance (also called the Earth Mover’s distance) quantifies the minimum cost required to transform one probability distribution into another. It is computed as the minimum of $c(x, \pi(x)) = d(x, \pi(x))^p$, where $d(x, \pi(x))$ represents the distance between $x$ and its transported counterpart under $\pi$. To compute this distance efficiently, the distributions are embedded onto a unit hypersphere using volumetric harmonic maps, $\phi_1: M \rightarrow S^n$ and $\phi_2: N \rightarrow S^n$, where $S^n$ denotes a canonical parameterized space in Cartesian coordinates. The empirical measures $X^\prime(x, \mu)$ and $Y^\prime(y, v)$ are defined on the transformed supports $\phi_1(\Omega_X)$ and $\phi_2(\Omega_Y)$, respectively. The optimal transport map $T: X^\prime \rightarrow Y^\prime$ induces the two-dimensional Wasserstein distance $W_2(X^\prime, Y^\prime)$, which is discretized over a finite point set $P(p, v)$ on the target domain. The squared Euclidean distance $||m_i - p_j||^2$ serves as the local transport cost between each pair of source and target points.

In our implementation, we adopt the discrete variational OT formulation described by~\citep{tu2020computing}, which models the mapping as a convex optimization problem. The algorithm iteratively updates a set of vertex weights that define a convex cell decomposition (also known as a lifted power diagram~\citep{tu2020computing,Gu2013}) until the transported mass between neighboring regions matches their target areas. The gradient of the transport energy guides the adjustment of these weights, while the Hessian matrix of the energy function ensures convergence through Newton’s method~\citep{villani2008optimal}. This iterative process continues until the difference between the source and target areas falls below a small threshold, yielding an area-preserving mapping between the two surfaces~\citep{tu2020computing}.

Within our pipeline, the OT-based mapping establishes an optimal correspondence between the original cortical surface and a planar parameterization, ensuring that the local area of each surface element is preserved after flattening. Unlike conformal mappings, which preserve angles but distort areas, the OT mapping minimizes local area distortion while maintaining global topological consistency. The resulting two-dimensional disk provides a smooth, bijective parameter domain where each cortical vertex corresponds uniquely to a planar vertex, forming a robust foundation for subsequent topological smoothing and CMF computation.

\subsubsection{Topological Smoothing}

Topological smoothing~\citep{Tu:Plos2021,tu2022protocol} is an iterative optimization procedure designed to smooth retinotopic maps while preserving their topological structure. The cortical surface is represented as a triangular mesh, and each geodesic disk patch is parameterized onto a two-dimensional unit disk using prior area-preserving optimal transportation method. Within this parametric domain, the retinotopic coordinates (eccentricity and polar angle) are iteratively refined under quasiconformal constraints, measured by the \textbf{Beltrami coefficient}~$\mu$, which must satisfy $\lvert \mu \rvert < 1$ for all triangles to ensure a diffeomorphic mapping~\citep{Tu:Plos2021}.

At each iteration, a new smooth map $\hat{v}$ is generated through three main steps:
\begin{enumerate}
    \item \textbf{Boundary update:} Boundary visual coordinates are fitted by local polynomial regression and restricted within a tolerance $t$ to avoid excessive displacement.
    \item \textbf{Topological correction:} Local orientation flips are detected using the Beltrami coefficient and corrected by averaging boundary vertices within $k$-nearest neighbors. The smoothing strength is adaptively increased (i.e., a larger $k$) until all triangles satisfy the topological constraint.
    \item \textbf{Laplacian smoothing:} The interior vertices are updated by solving the following linear system~\citep{Eilers2003,Garcia2010,Tu:Plos2021}:
    \[
    (\mathrm{diag}(R^2) + \lambda \Delta)\hat{v} = \mathrm{diag}(R^2)v,
    \]
    where $\Delta$ denotes the discrete Laplace-Beltrami operator, $R^2$ denotes the pRF-decoding reliability, and $\lambda$ denotes the smoothness factor.
\end{enumerate}

The iterations continue until convergence, defined by both the topological condition ($\lvert \mu \rvert < 1$) and the stability of vertex updates bounded by the parameter $\textit{meanddth}$. In addition, the loop is limited to a maximum of 20 iterations to prevent unbounded smoothing. Our parameter choices follow the protocol described in~\citep{tu2022protocol} but with stricter conditions: $t=0.05$, $\lambda=0.001$, $k=3$, and $\textit{meanddth}=1.0$. In practice, most subjects satisfied these criteria well before the 20-iteration cap. The resulting map achieves a strong balance between surface smoothness and faithful preservation of cortical topology.

In application to the 181 HCP subjects, the algorithm successfully converged for all cases within 20 iterations. After smoothing, every subject’s retinotopic map became fully topologically correct, exhibiting zero flipped triangles. The process maintained diffeomorphic consistency while preserving close agreement with the original pRF decoding results, showing only a minor reduction in the explained variance ($R^2$) due to the topological constraint~\citep{Tu:Plos2021}. Importantly, the method preserves genuine biological discontinuities—such as those potentially arising in clinical or pathological conditions—because topological correction only targets spurious local flips while maintaining the global organization of the map~\citep{Tu:Plos2021}.

\begin{figure}[t]
\begin{center}
\begin{tabular}{c}
\includegraphics[width=0.975\linewidth]{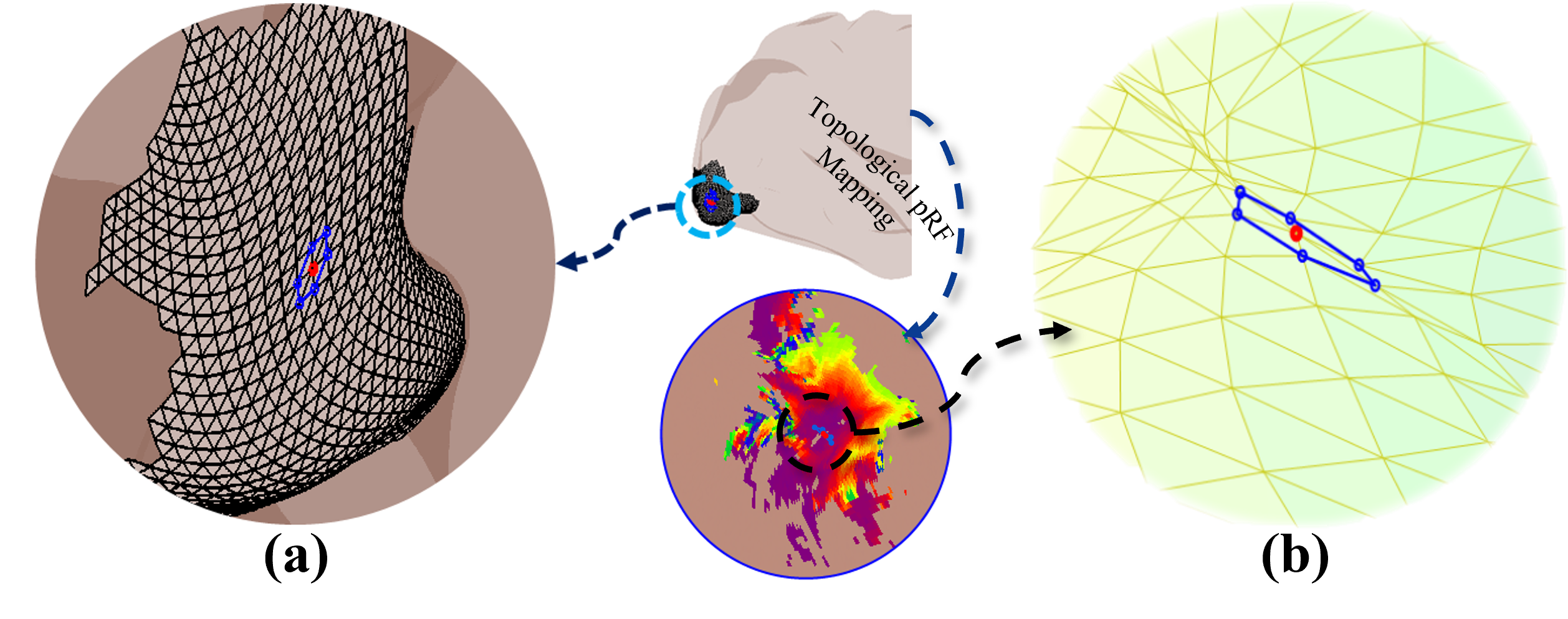}
\end{tabular}
\end{center}
\caption 
{ \label{fig:2}
1-ring patch measurement: estimation of cortical magnification factor by the division between areas of the dual rings. (a) A vertex’s dual ring structure on 3D brain mesh, (b) The dual ring structure on the retina visual space.} 
\end{figure}

\subsubsection{One-Ring Patch Method}
The Cortical Magnification Factor (CMF) can be estimated using a local structural method based on the vertex dual structure over a one-ring patch. The vertex one-ring is defined as the set of immediate neighbors of a center vertex, i.e., all vertices that can be reached by traversing a single edge. Fig.~\ref{fig:2}a illustrates a vertex (the central red point) and its associated one-ring patch, represented by the enclosed blue polygon.

After the cortical surface within the V1–V3 complex is topologically smoothed, each vertex on the three-dimensional mesh is assigned a receptive-field location in visual space. This provides a topologically correct retinotopic map in which each cortical vertex corresponds to a unique visual coordinate. For a given vertex $v_i$, we identify its one-ring neighborhood $\mathcal{N}(v_i)$ on the 3D mesh and compute two quantities: (1) the cortical-surface area, defined as the sum of the face areas enclosed by $\mathcal{N}(v_i)$; and (2) the visual-field area, defined as the polygonal area formed by the mapped receptive field centers $\{v'\}$ in visual space corresponds to $\mathcal{N}(v_i)$ (Fig.~\ref{fig:2}b). The CMF at vertex $v_i$ is then given by the ratio of these two areas:  
\[
\text{CMF}(v_i) = \frac{\text{Area}_{\text{cortex}}}{\text{Area}_{\text{visual}}} = \frac{\text{Area}_{\mathcal{N}(v_i)}}{\text{Area}_{\{v'\}}}
\]





Once the smoothed retinotopic map is obtained, the final CMF estimation is carried out entirely on the 3D surface, consistent with prior approaches~\citep{schira2007two, schira2009foveal, kupers2021population, kupers2022asymmetries}. However, this one-ring patch method computes an areal CMF (cortical surface area per unit visual field area), whereas prior studies report linear cortical magnification (cortical distance per degree of visual angle). Consequently, the numerical CMF values obtained here are not directly comparable to those of ~\citep{schira2009foveal, kupers2022asymmetries}. Nonetheless, both methods reveal qualitatively similar patterns of visual field organization, with the highest magnification at the fovea and a decline with increasing eccentricity. This method relies on topologically accurate retinotopic mapping: non-topological maps can lead to inverted or flipped neighborhoods ${v'}$ in visual space, generating highly noisy and unreliable CMF estimates. By incorporating topological constraints via optimal transport flattening and Beltrami coefficient–based smoothing, our pipeline ensures robust and physiologically consistent results.

This method relies on topologically accurate retinotopic mapping: non-topological maps can lead to inversed or flipped neighborhood $\{v'\}$ in visual space, generating highly noisy and unreliable CMF estimates. By incorporating topological constraints via optimal transport flattening and Beltrami coefficient–based smoothing, our pipeline ensures robust and physiologically consistent results.

Finally, the CMF results from the left and right hemispheres were merged to provide a unified representation of the perceptual resource distribution across the visual field.

\subsubsection{Merge Per-Hemisphere CMF to Entire Visual Field}
\begin{figure}[t]
\begin{center}
\begin{tabular}{c}
\includegraphics[width=0.85\linewidth]{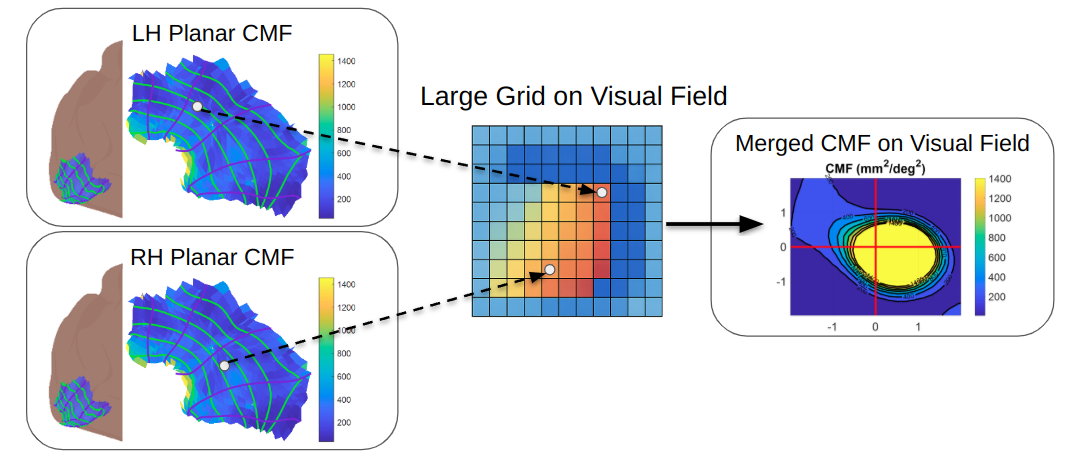}
\end{tabular}
\end{center}
\caption 
{ \label{fig:merge}
Illustration of the merging method: CMF results from left (LH) and right (RH) hemispheres are projected onto a common visual field grid and combined into a unified CMF representation.}
\end{figure}

Because fMRI measurements provide cortical surface reconstructions for the left and right hemispheres separately, the $\text{CMF}(v_i)$ values were first computed for each hemisphere independently. To obtain a unified representation of cortical magnification across the entire visual field and to illustrate individual patterns of visual concentration, the results from the two hemispheric  were merged onto a common two-dimensional coordinate system, as shown in Fig.~\ref{fig:merge}.

Specifically, the CMF values for each hemisphere were projected from their cortical surface vertices to the corresponding receptive-field centers in visual-field coordinates $(\rho, \theta)$, where $\rho$ denotes eccentricity and $\theta$ denotes polar angle. A uniform $100 \times 100$ grid was defined over the visual field to serve as the reference domain. For each grid location, CMF values were interpolated from the nearest cortical-vertex measurements, with separate passes for the left and right hemispheres. Hemispheric assignments were restricted to their valid angular ranges (right hemisphere: $-0.5\pi < \theta < 0.5\pi$; left hemisphere: $0.5\pi < \theta < 1.5\pi$ with sub-ranges shifted by $\pi$ to account for mirror symmetry).

The CMF values from both hemispheres were aggregated within the grid cells. When multiple vertices contributed to the same grid cell, their CMF values were averaged to yield a single representative estimate. This procedure was carried out both in polar coordinates $(\rho, \theta)$ and, after coordinate conversion, in Cartesian coordinates $(x,y)$. This enabled consistent visualization of cortical magnification across the entire visual field. As a quality check, we confirmed that nearly all grid points within the defined field of view were assigned valid CMF values after merging, indicating successful reconstruction of complete visual-field magnification maps. 

This per-hemisphere merging procedure ensured that cortical magnification was computed consistently across subjects and that both hemispheres were seamlessly integrated into a unified, whole-field planar representation suitable for further group-level analysis.

\section{Results}

\subsection{Optimal Transportation and Area Preserving}


\begin{figure}[t]
\begin{center}
\begin{tabular}{c}
\includegraphics[width=0.55\linewidth]{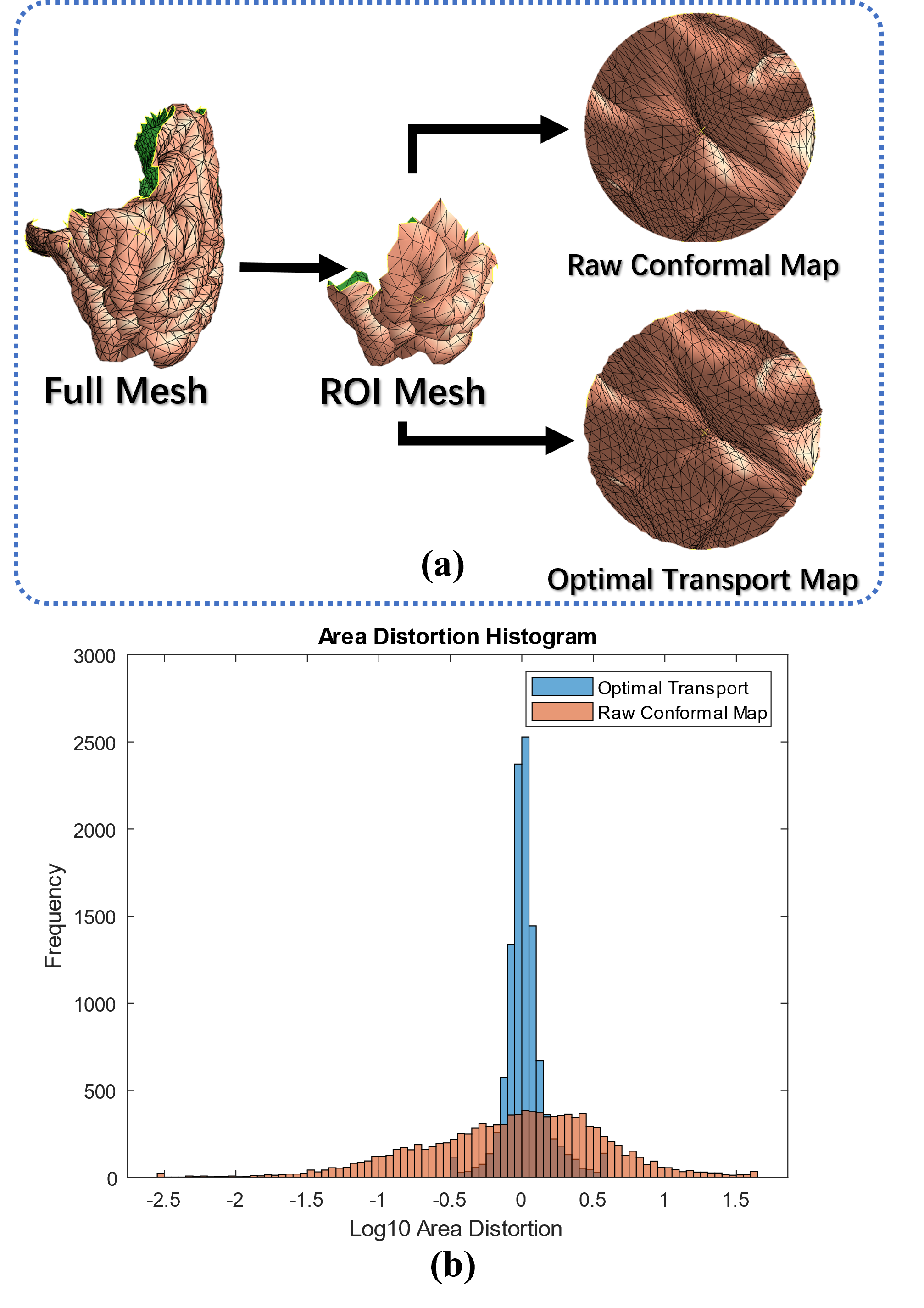}
\end{tabular}
\end{center}
\vspace*{-7mm}
\caption
{(a) Comparison of raw conformal mapping and optimal transport mapping from 3D brain mesh to 2D unit disk. (b) Comparison of area distortion between these 2 methods.}\label{fig:area}
\end{figure}

In our pipeline, optimal transport was used to compute an optimal area-preserving mapping between the original cortical surface and a parameterized planar disk. To evaluate its effectiveness, we conducted an ablation study on all 181 subjects in the HCP dataset. We quantified area distortion across all cortical mesh triangles during the transformation from the 3D brain structure to the 2D planar surface. The quality of this flattening---and its suitability for CMF measurements---was assessed by measuring the area distortion at each discrete triangle, comparing the original 3D mesh with the resulting flattened surface~\citep{Ta2022,su2015optimal}. 

Conventional conformal mapping, which corresponds to smooth complex analytic functions, minimizes angle distortion but introduces significant area distortion. This trade-off undermines its suitability for 2D CMF measurements using area-based methods, such as the 1-ring patch approach. In contrast, integrating OT into the mapping process substantially reduced area distortion.

As shown in Fig.~\ref{fig:area}, the incorporation of OT drastically reduced area distortion across the cortical mesh, approaching near-zero values, particularly within regions of interest (ROI). This enhanced area preservation enables more accurate CMF estimation through the 1-ring patch method, which relies on precision area measurements between the visual field and the cortical surface.

\subsection{CMF from the HCP Dataset}
The retinotopy data from all 181 healthy participants in the Human Connectome Project (HCP)\citep{benson2018hcp} where high quality 7T retinotopy data was acquired, were processed in this study using the proposed pipeline. After preprocessing of the entire early visual cortex, we measure CMF solely on V1 and reported the results as follows.


\subsubsection{CMF Results and Individual Differences}
To analyze individual differences in CMF patterns across subjects, we applied K-means clustering to the CMF maps derived from the 181 observers in the HCP dataset. Specifically, the raw 2D matrices of CMF values in visual-field coordinates were used directly as input to the clustering, without additional feature extraction or dimensionality reduction (e.g., PCA). The optimal number of clusters was determined to be $K = 5$, as indicated by the elbow point in the loss function curve (Fig.~\ref{fig:HCPElbowResults}), reflecting the most significant reduction in loss without overfitting. To ensure robustness, the clustering procedure was repeated across multiple random initializations, and the resulting solutions consistently converged to similar cluster assignments.

\begin{figure}[ht]
\begin{center}
\begin{tabular}{c}
\includegraphics[width=0.97\linewidth]{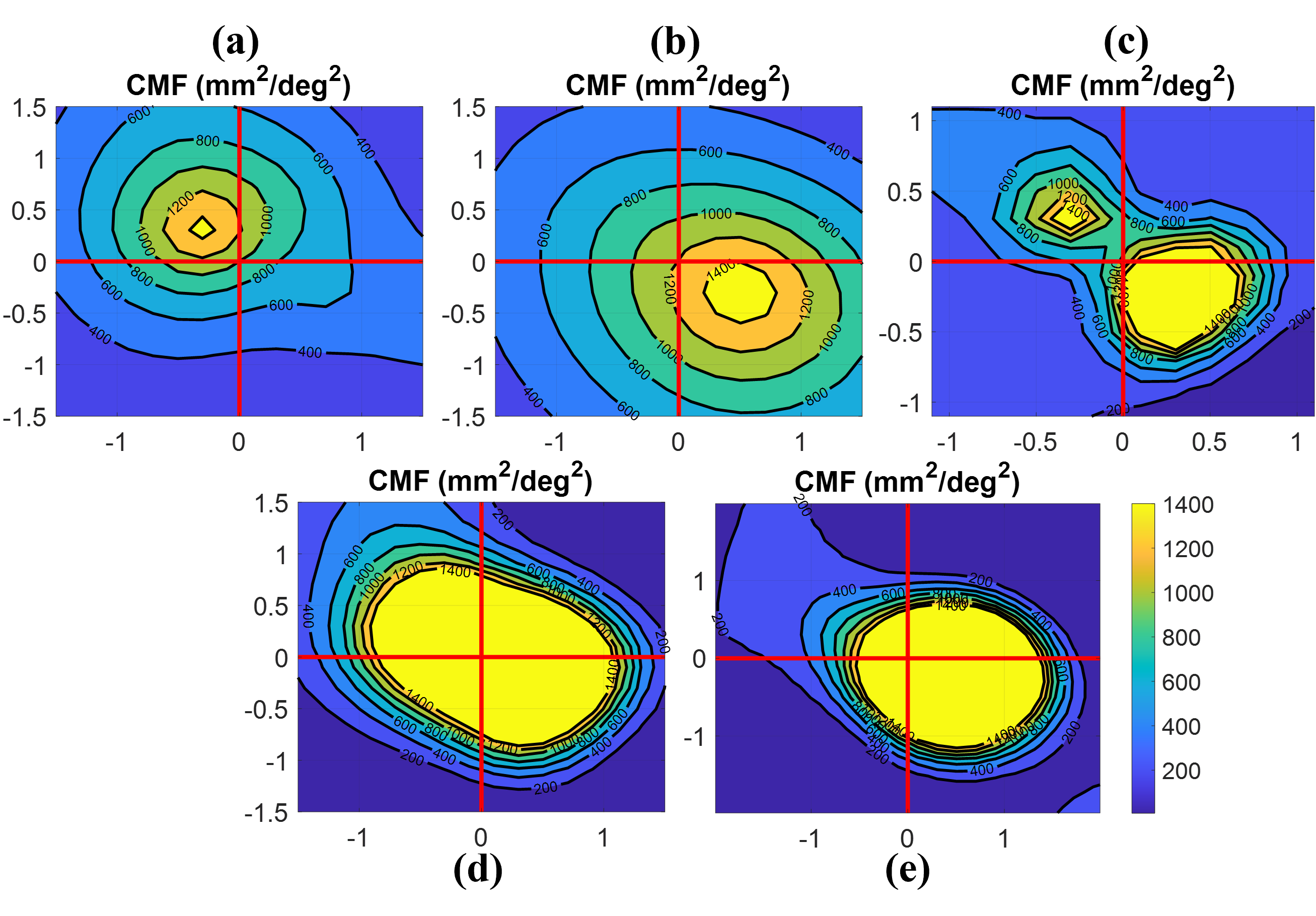}
\end{tabular}
\end{center}
\vspace*{-7mm}
\caption 
{ \label{fig:CMF-OT}
 Average 2D planar V1 CMF results for 5 different clusters in the HCP cohort: (a) Upper-Left Cluster, CMF concentrated on the upper-left of the fovea in the visual field. (b) Lower-Right Cluster, CMF concentrated on the lower-right of the fovea. (c) Dual-Centralized Cluster, CMF held two concentration points on the upper-left and lower-right, respectively. (d) Centralized Cluster, CMF generally concentrates on the center of the fovea. (e) Lower-Centralized Cluster, CMF concentrates on the center with a slight offset to the bottom.} 
\end{figure}

This clustering allowed us to categorize subjects into groups with similar CMF profiles, revealing distinct patterns of visual information processing throughout the population. The average 2D planar CMF results for each cluster are shown in Fig.~\ref{fig:CMF-OT}.

\begin{figure}[t]
\begin{center}
\begin{tabular}{c}
\includegraphics[width=0.75\linewidth]{Fig7_kmeans_HCP.png}
\end{tabular}
\end{center}
\vspace*{-7mm}
\caption
{Elbow plot for K-means clustering of HCP CMF results. The plot shows the inertia (within-cluster sum of squares) as a function of the number of clusters. The optimal number of clusters, identified at the elbow point, is K = 5 when aggregating CMF data from all 181 HCP subjects.}\label{fig:HCPElbowResults}
\end{figure}

In all subjects, the CMF center was predominantly located near the foveal region, consistent with previous findings~\citep{benson2021cortical, himmelberg2021cross, harvey2011relationship, silva2018radial, kupers2022asymmetries, himmelberg2022linking}. In quantitative terms, our estimated CMF magnitudes were higher than those reported in earlier studies, owing largely to differences in measurement approach. For example, \citep{schira2009foveal, kupers2021population, kupers2022asymmetries} observed linear cortical magnification factors on the order of $2\sim4 \times$ at approximately 5$^\circ$ eccentricity (roughly 2–4 mm of V1 per degree of visual angle), whereas our one-ring areal method yielded values around $10\sim15 \times$ at the same eccentricity. These larger magnitudes are expected, because areal CMF inherently scales with the square of linear magnification, and our approach (which averages over a 1-ring patch and applies smoothing) produces a flatter profile, slightly reducing peak foveal values while elevating magnification at moderate eccentricities. Importantly, despite the numeric discrepancies, both our findings and those of Schira and Kupers demonstrate the same magnification trends and visual-field structure: cortical magnification declines with increasing eccentricity and exhibits comparable anisotropies across the visual field. Thus, the differences in CMF magnitude reflect methodological differences (areal vs. linear, smoothing and averaging techniques) rather than contradictory results in cortical organization.

\begin{table}[t]
\centering
\caption{Number of observers for each cluster in the HCP cohort.}\label{tab:HCPHeadcount}
\begin{tabular}{|l|c|}
\hline
Clusters & \# of observers \\
\hline
a. upper-left & 32 \\
\hline
b. lower-right & 26 \\
\hline
c. dual-centralized & 11 \\
\hline
d. centralized & 47 \\
\hline
e. lower-centralized & 68 \\
\hline
\end{tabular}

\end{table}

Furthermore, our clustering analysis revealed clear inter-subject variability in CMF asymmetry and concentration behavior, especially along the polar angle dimension. Table~\ref{tab:HCPHeadcount} summarizes the number of subjects assigned to each cluster. Most of the participants (115 of 181) fell into the centralized (47 subjects) or lower centralized (68 subjects) clusters, exhibiting a broad concentration of CMF that declined radially outward from the fovea. Despite this similarity, the two clusters differed in symmetry: the lower-centralized group showed a downward asymmetry, while the centralized group displayed a more symmetric distribution.

The remaining 66 subjects showed more localized and quadrant-specific patterns. Of these, 32 exhibited CMF concentration in the upper-left quadrant (Fig.~\ref{fig:CMF-OT}b), and 11 subjects showed a dual centralized configuration with two distinct focal points (Fig.~\ref{fig:CMF-OT}c), suggesting an unusual but consistent bilateral pattern.

To the best of our knowledge, no previous studies have systematically characterized this level of individual variability in CMF concentration behavior across the visual field. By categorizing participants into five distinct CMF profiles, our findings provide novel insights into how individuals may differ in their visual field processing. These differences, which are difficult to articulate through verbal descriptions alone, underscore the need for more research into the neural and functional implications of this diversity of the CMF.

\subsubsection{Correlation between CMF and Receptive Field Size}
\begin{figure}[t]
\begin{center}
\begin{tabular}{c}
\includegraphics[width=0.95\linewidth]{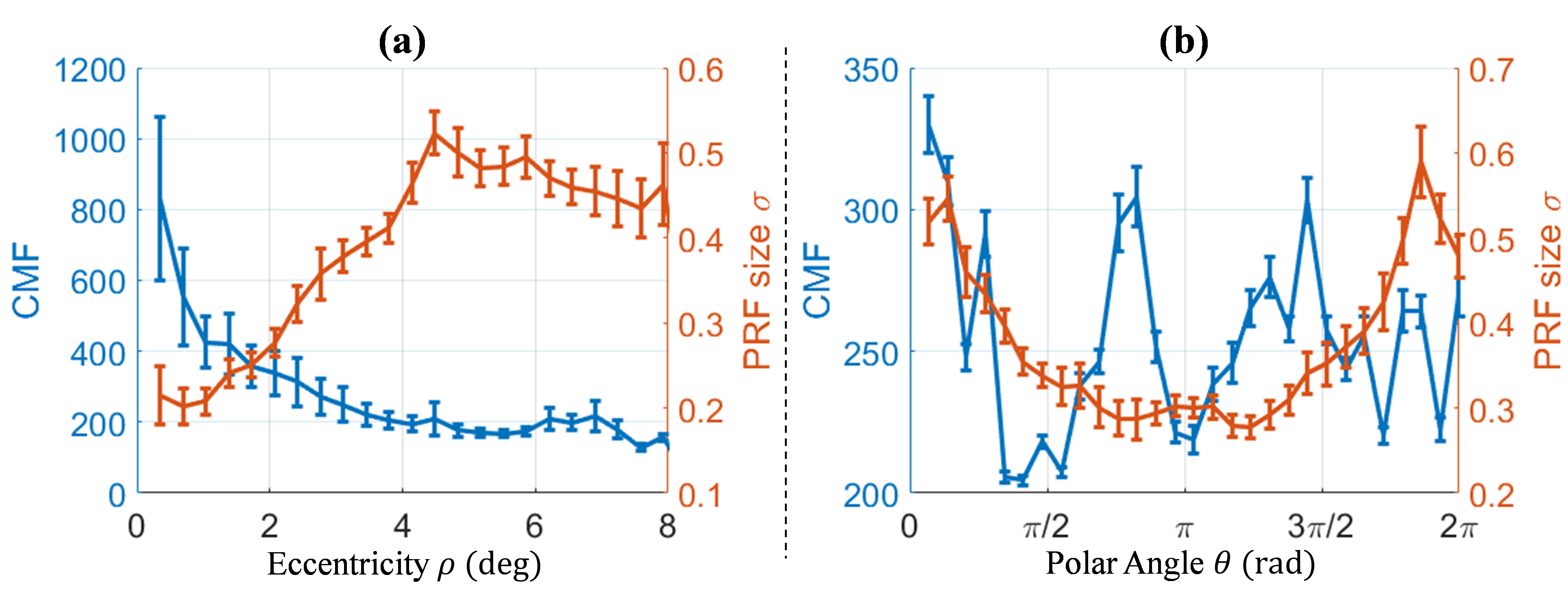}
\end{tabular}
\end{center}
\vspace*{-7mm}
\caption 
{ \label{fig:41}
 Correlation between CMF and pRF size across (a) eccentricity $\rho$ and (b) polar angle $\theta$ } 
\end{figure}

In this study, we also examined the characteristics of pRF sizes across the visual field. Our results were consistent with previous literature~\citep{Adams2003,van1984visual,harvey2011relationship,Hubel1974,silva2018radial,kupers2022asymmetries} and revealed asymmetries in pRF size with respect to polar angle that aligned with the results of our planar CMF measurements. Specifically, we found a negative correlation between the pRF size $\sigma$ and CMF: as points moved outwards from the fovea (i.e., with increasing eccentricity), the pRF size $\sigma$ increased while the CMF decreased (Fig.~\ref{fig:41}a). We also observed that pRF size $\sigma$ and CMF were evenly distributed across different polar angles (Fig.~\ref{fig:41}b). These findings align with prior studies~\citep{benson2021cortical, silva2018radial, kupers2022asymmetries, himmelberg2022linking}, which report that visual acuity is highest near the fovea and declines with eccentricity.

To provide quantitative support for these observations, we computed Pearson correlation coefficients and associated p-values across the 181 HCP subjects. The results are summarized in Table~\ref{tab:CMF_pRF_corr}. A strong positive correlation was observed between pRF size and eccentricity, while CMF showed a strong negative correlation with eccentricity. Importantly, pRF size and CMF were negatively correlated, confirming the inverse relationship between these two measures.

\begin{table}[thbp]
\centering
\caption{Quantitative correlations between pRF size, CMF, and eccentricity across HCP subjects.}
\label{tab:CMF_pRF_corr}
\begin{tabular}{lcc}
\hline
Variables & Pearson $R$ & $p$-value \\
\hline
pRF size vs. Eccentricity & 0.749 & 1.643$\times 10^{-5}$ \\
CMF vs. Eccentricity & -0.770 & 6.821$\times 10^{-6}$ \\
pRF size vs. CMF & -0.695 & 1.140$\times 10^{-4}$ \\
\hline
\end{tabular}
\end{table}

It has been widely accepted that pRF size and CMF exhibit a negative correlation across variables such as eccentricity and polar angle~\citep{harvey2011relationship}. However, our methodology allowed for a novel approach by projecting decoded pRF results directly onto the 2D visual field, enabling precise calculation and visualization of planar CMF. This projection made it possible to directly examine the correlation between pRF responses and CMF within the 2D planar visual domain.

\begin{figure}[htpb]
\begin{center}
\begin{tabular}{c}
\includegraphics[width=0.70\linewidth]{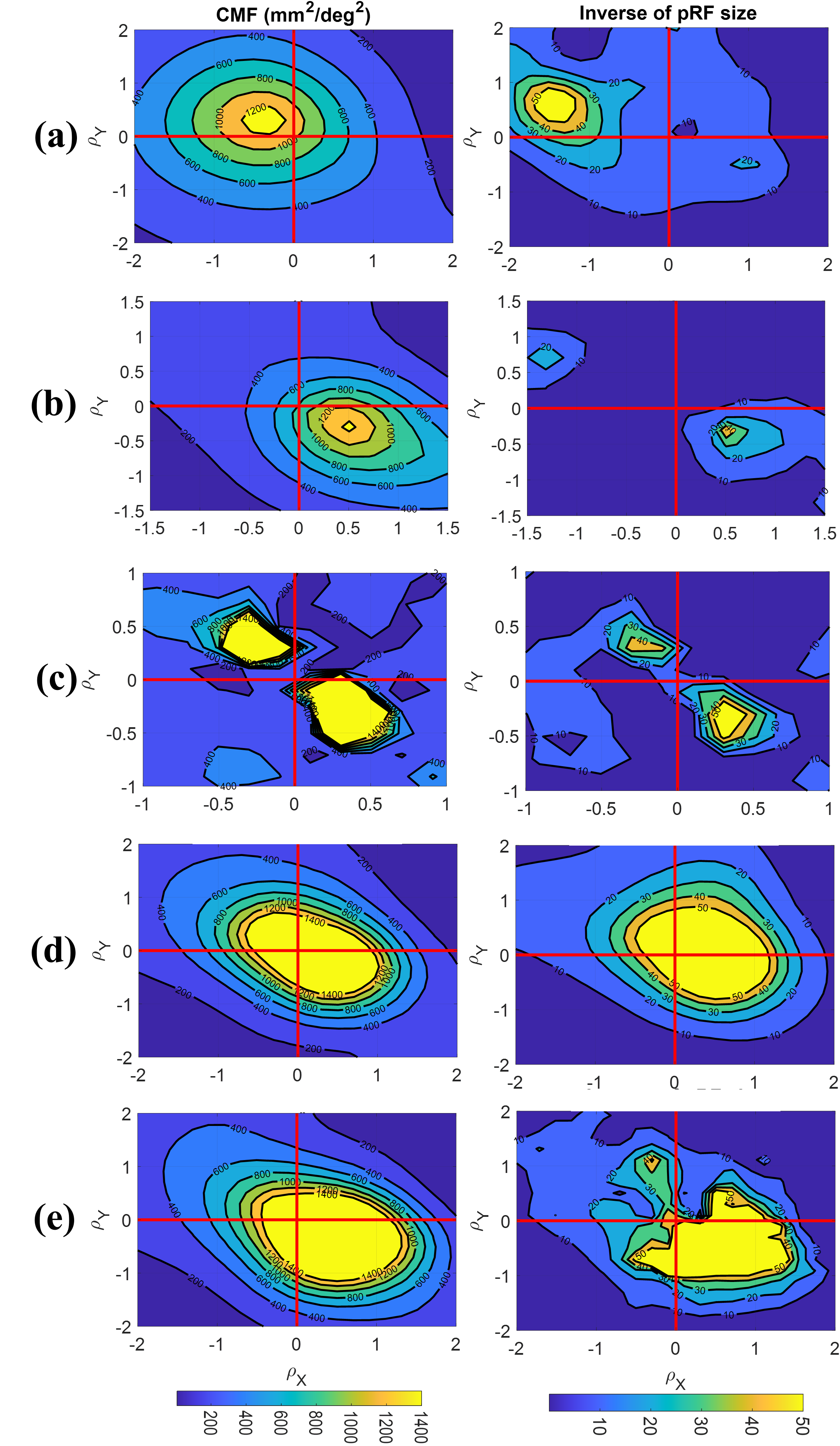}
\end{tabular}
\end{center}
\vspace*{-7mm}
\caption 
{ \label{fig:5}
Example comparisons of planar CMF and the inverse of pRF size, $1/\sigma$: (a) subject 104416 (upper-left), (b) subject 114823 (lower-right), (c) subject 905147 (dual-centralized), (d) subject 146129 (centralized), and (e) subject 126426 (lower-centralized).}
\end{figure}

In addition to characterizing CMF patterns, we analyzed pRF solutions across all observers, as illustrated in Fig.~\ref{fig:5}. Notably, the inverse of the pRF size $1/\sigma$ was found to closely align with our planar CMF measurements, further supporting the negative relationship between pRF size and cortical magnification in the visual field.

It is well established that smaller pRF sizes or larger CMF values indicate higher visual acuity near a given point, while acuity tends to decrease with increasing eccentricity. Consequently, regions farther from the fovea usually exhibit larger pRF sizes and smaller CMF values. In our study, we extended this understanding by analyzing the behavior of the pRF size in planar coordinates, finding that pRF size not only increased with eccentricity but also varied systematically with polar angle.

Furthermore, we observed that, for most subjects, the inverse of the pRF size $1/\sigma$ closely aligned with their CMF-based cluster. For instance, subject 905147 displayed a dual-centralized pattern in both their planar CMF map and their $1/\sigma$ distribution. This consistency between CMF and pRF measurements supports the validity of our clustering approach and reinforces the assumed correlation between the two metrics.

Collectively, these findings demonstrate that individual differences in visual field representation can be robustly characterized using both CMF and pRF analyses, offering complementary insights into perceptual variability across subjects. The addition of quantitative correlation metrics strengthens the evidence for the negative relationship between pRF size and CMF.

\subsubsection{CMF Asymmetries}

In this study, we measured the planar CMF for all 181 observers in the HCP cohort on their V1 cortex. Consistent with previous research~\citep{Adams2003,van1984visual,harvey2011relationship,Hubel1974,silva2018radial,kupers2022asymmetries}, our observations revealed asymmetric visual field representations near the fovea. Specifically, some individuals exhibited increased CMF in the upper visual field, while others showed stronger representation in the lower visual field. Our planar CMF clustering analysis captured these individual differences, revealing distinct CMF behaviors around the fovea region. 

\begin{figure*}[t]
\begin{center}
\begin{tabular}{c}
\includegraphics[width=0.95\linewidth]{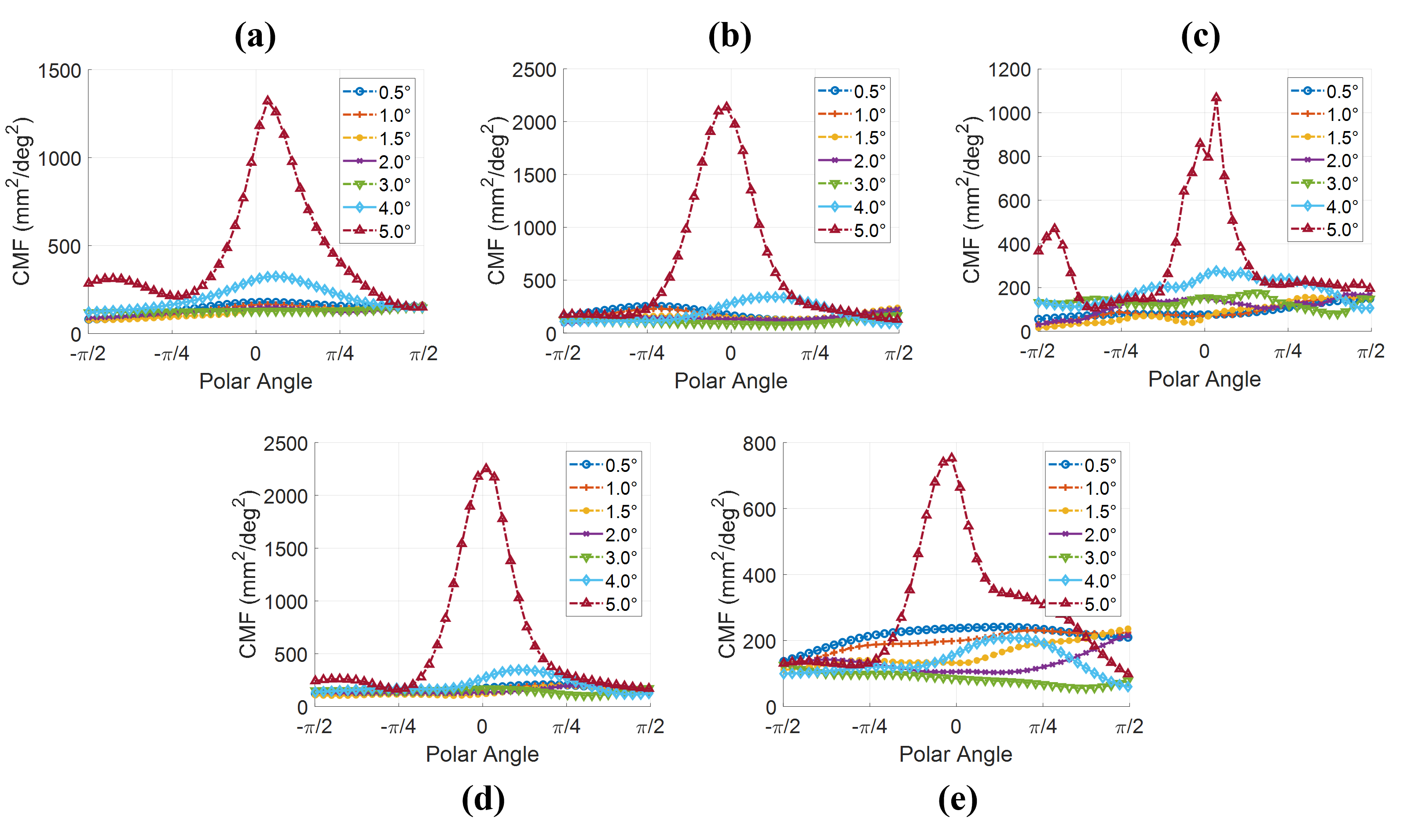}
\end{tabular}
\end{center}
\vspace*{-7mm}
\caption 
{Examples of visual asymmetry in CMF as a function of polar angle across several eccentricities: (a) subject 104416 (upper-left), (b) subject 114823 (lower-right), (c) subject 905147 (dual-centralized), (d) subject 146129 (centralized), and (e) subject 126426 (lower-centralized) } \label{fig:asymmetry}
\end{figure*}

Unlike previous approaches that assessed asymmetry by comparing broad horizontal (left vs. right) or vertical (upper vs. lower) quadrants~\citep{silva2018radial, harvey2011relationship}, our method used planar CMF maps in two dimensions, incorporating both eccentricity and polar angle. This enabled a more nuanced assessment of visual field organization, revealing subject-specific asymmetry patterns near the fovea (Fig.~\ref{fig:asymmetry}). For example, observers with dominant CMF concentration in the upper-left versus lower-right visual field exhibited opposing asymmetry patterns, whereas those with a centralized CMF distribution displayed near-radial symmetry around the fovea.

In summary, our findings confirmed that early human visual field representations—especially in areas V1—were not radially symmetrical in terms of either CMF or pRF size. While the inverse relationship between pRF size and CMF with increasing eccentricity was consistent with classical models~\citep{Hubel1974}, our results revealed that the location and extent of CMF asymmetry can vary considerably across individuals~\citep{harvey2011relationship}. This analysis provided a more nuanced understanding of individual differences in visual perception, highlighting the role of radial asymmetries in early visual processing.

\subsection{CMF Validation Using the NYU Retinotopy Dataset}

\begin{figure*}[t]
\begin{center}
\begin{tabular}{c}
\includegraphics[width=0.75\linewidth]{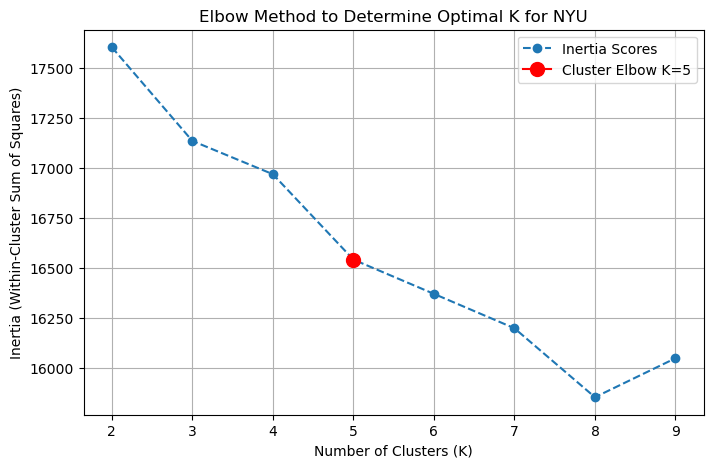}
\end{tabular}
\end{center}
\vspace*{-7mm}
\caption
{Elbow plot for K-means clustering of NYU CMF results. The plot shows the inertia (within-cluster sum of squares) as a function of the number of clusters. The optimal number of clusters, identified at the elbow point, is K = 5 when aggregating CMF data from all 42 NYU subjects.}\label{fig:NYUElbowResults}
\end{figure*}

After analyzing CMF measurements from 181 observers in the HCP dataset, it was essential to validate our findings using an independent dataset with different experimental parameters and participant demographics. This step was critical in ensuring the robustness and generalizability of our observations on CMF asymmetries and clustering patterns. Although the HCP dataset offered high-resolution data acquired under specialized conditions, including 7T MRI scanners, long-duration scans, and a large cohort, it also reflected a unique experimental design. Validating our findings with alternative datasets helps control for possible biases or artifacts related to the HCP's specific protocol.

Recent studies~\citep{himmelberg2021cross, allen2022massive, gong2023large} have shown strong agreement with HCP on key retinotopic properties, despite differing setups. Thus, these independent datasets serve as important benchmarks for testing the reproducibility of retinotopic mapping and CMF estimation. In particular, they offer variability in scanner strength, task design, and subject cohorts, factors essential to confirm whether our CMF-based clustering and asymmetry findings are independent of the data set.

Among the available alternatives, we selected the New York University (NYU) Retinotopy Dataset~\citep{uugurbil2013pushing, himmelberg2021cross}, 
which includes a relatively large sample (44 subjects)—more than other public datasets~\citep{allen2022massive, gong2023large}—making it a suitable cohort to validate the consistency of the behavioral clusters of CMF.

\begin{figure*}[t]
\begin{center}
\begin{tabular}{c}
\includegraphics[width=\linewidth]{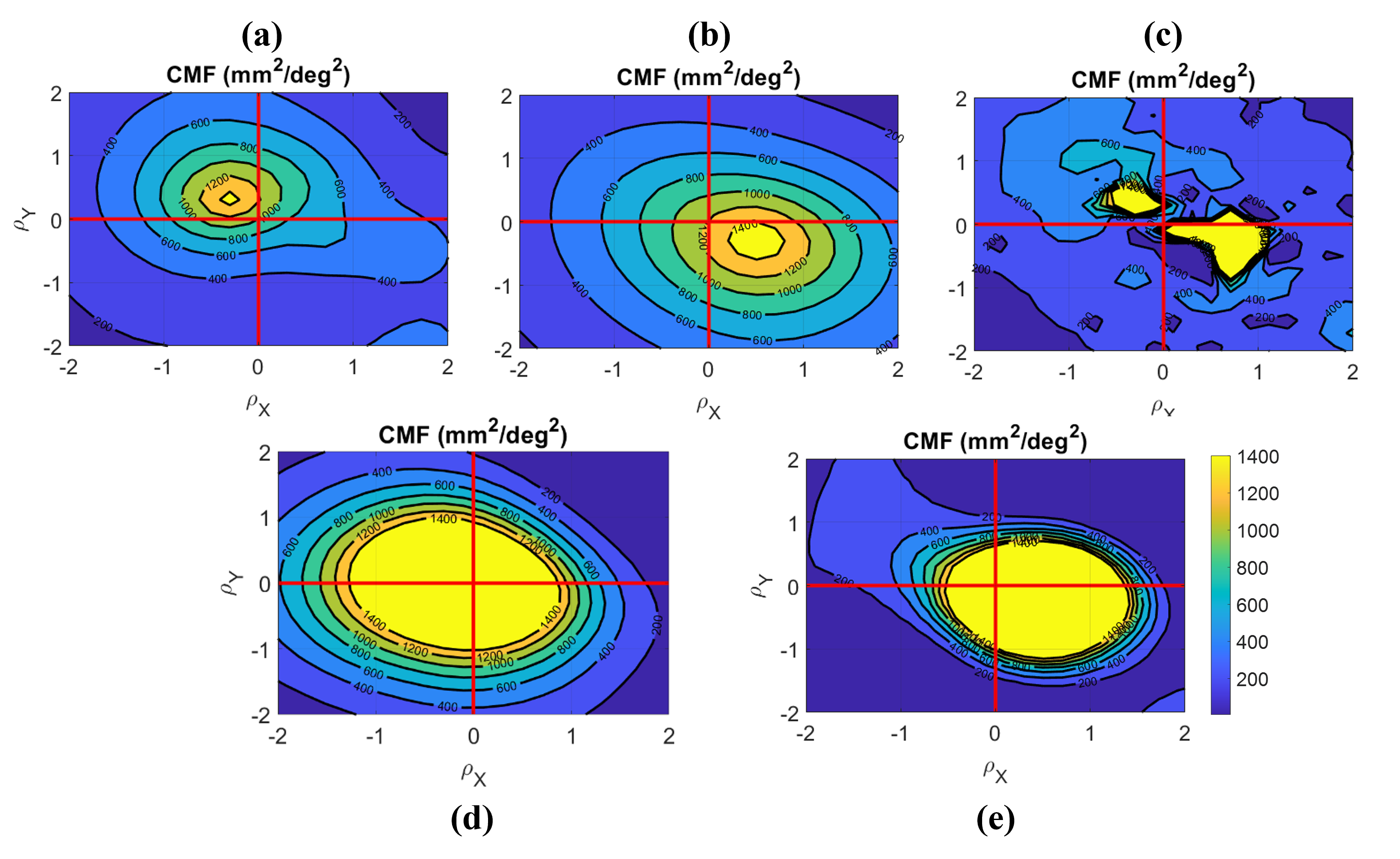}
\end{tabular}
\end{center}
\vspace*{-7mm}
\caption
{Average CMF results for all 5 clusters from the NYU dataset.}\label{fig:nyu_all}
\end{figure*}

We applied our CMF estimation pipeline to 42 of the 44 NYU subjects and performed K-means clustering on the resulting CMF maps. Fig.~\ref{fig:NYUElbowResults} illustrates the inertia (within-cluster sum of squares) as a function of the number of clusters. The optimal number of clusters, identified at the elbow point, is $K=5$ when aggregating the CMF data from the 42 subjects of NYU.  Fig.~\ref{fig:nyu_all} further shows the average CMF results of the five groups. They were consistent with those identified in the HCP dataset. Table~\ref{table:nyu3T} summarizes the number of subjects in each NYU cluster, highlighting the persistence of CMF patterns in different datasets.

\begin{table}[ht]
\centering
\caption{Number of subjects for each cluster in the NYU retinotopy dataset.}
\begin{tabular}{|l|c|}
\hline
Clusters &  \# of subjects in NYU 3T Dataset\\
\hline
a. upper-left & 8 \\
\hline
b. lower-right & 14 \\
\hline
c. dual-centralized & 12 \\
\hline
d. centralized & 2\\
\hline
e. lower-centralized & 6\\
\hline
\end{tabular}
\label{table:nyu3T}
\end{table}

This cross-dataset validation reinforced the reliability of our CMF estimation and clustering methods. The recurrence of consistent CMF behavior patterns across both high-field (HCP) and standard-field (NYU) datasets confirmed that our findings were not artifacts of a specific imaging protocol or participant group. These results support the broad applicability of our methods to various visual neuroscience studies and strengthen the case for using CMF-based analysis to capture meaningful individual differences in early visual cortex organization.

\section{Discussion}

\subsection{Limitations of Traditional CMF Quantification}

Current methods for quantifying the Cortical Magnification Factor (CMF) exhibit several limitations. Despite their widespread adoption~\citep{horton1991representation, virsu1996cortical, duncan2003cortical, song2015neural, hussain2015estimation, bordier2015quantitative, lage2020investigating, benson2021cortical, himmelberg2021cross, himmelberg2023polar}, conventional approaches often lack precision and scalability. Key challenges include:

\begin{itemize}
\item Manual ROI Definition: Many studies rely on hand-drawn contours for defining regions of interest (ROIs), which is time-consuming, lacks standardized criteria, and impairs reproducibility.
\item 1D CMF Measures: Most methods compute CMF linearly along eccentricity axes, neglecting the inherently 2D nature of retinotopic representation.
\item Geometric Inaccuracies: Existing projection techniques (e.g., graph distances, orthographic projections) distort the geometry of cortical surfaces and do not preserve topological continuity.
\end{itemize}

These limitations can introduce distortions in cortical area estimation, reduce measurement accuracy, and impair continuity across the visual field, ultimately hampering a comprehensive understanding of cortical magnification.

\subsection{Advancing CMF Estimation with Optimal Transport}
To address these issues, we introduced an optimal transport (OT) framework for area-preserving cortical surface parameterization~\citep{su2015optimal}. Building on quasiconformal geometry, our method incorporates the Beltrami coefficient (BC) to enforce topology-preserving and smooth mappings~\citep{Tu:Plos2021}. This framework supports the accurate computation of 2D areal CMF using area ratios of 1-ring patches and can extend to multiple visual areas (V1, V2, V3) through linear transformations of polar angle maps.

Our topology-preserving smoothing and OT-based parameterization significantly improved retinotopic mapping. The BC helped maintain continuity and minimize distortion, while additional phase correction steps resolved artifacts such as phase jumps. Together, these components facilitated precise localization of the receptive field, enabling robust planar CMF measurements.

We validated our method on both 7T HCP (181 subjects) and 3T NYU (44 subjects) datasets, demonstrating that our pipeline yielded consistent and reliable CMF estimates under varying experimental conditions. This highlights the potential of our framework as a versatile tool for retinotopic mapping in both research and clinical settings.

\subsection{Individual Differences in CMF}
Previous studies have reported perceptual asymmetries in cortical magnification, such as horizontal-vertical asymmetry (HVA)\citep{schwarzkopf2013subjective}.
These findings suggest that CMF is not a uniform function of eccentricity but reflects individual visual preferences and performance biases, reinforcing the need for personalized retinotopic analysis in both basic and translational research.

\subsection{Impact of Optimal Transport}
While conformal mapping methods~\citep{Tu:Plos2021,Ta2022,Tu:BSF2022} have previously succeeded in preserving angular information and smoothing retinotopic maps, they proved insufficient for accurate CMF evaluation. Specifically, we observed that the application of topology smoothing alone could generate abnormal CMF values near the fovea due to overlapping or degenerate vertex rings.
To mitigate this, we introduced optimal transport-based re-parameterization, which recalibrated visual coordinates and improved vertex spacing. As shown in Table~\ref{table4}, this adjustment reduced abnormal CMF points by over 66 percent, making the data suitable for downstream tasks such as clustering.

\begin{table}[t]
\centering
\caption{Comparison of parameterization methods before CMF measurement}
\label{table4}
\begin{tabular}{|l|c|}
\hline
Methods & \# of abnormal CMF points\\
\hline
1. Conformal Mapping + Topology smoothing & 267  \\
\hline
2. \textbf{Proposed} (OT + Topology smoohing) & 89 \\
\hline
\end{tabular}
\end{table}

\subsection{Limitations and Future Directions}
While our approach benefited from the high resolution and field strength of the 7T HCP dataset, limitations remained due to the inherent spatial resolution and signal-to-noise constraints in retinotopic imaging~\citep{benson2018hcp}.

Our study focused on the early visual areas and extending this work to higher-level visual areas (e.g., V4, MT) poses challenges due to reduced signal strength and increased variability in these regions. Additionally, clinical applications represent a promising future direction. Applying our framework to populations with visual disorders could improve understanding of functional reorganization and guide interventions for vision rehabilitation.

\section{Conclusion}

We developed a novel framework that integrates optimal transport and topological smoothing for accurate, area-preserving retinotopic mapping and 2D planar CMF estimation. Our method corrects topological violations and enhances the precision of CMF quantification, enabling robust comparisons between subjects and datasets.

Applying this framework to 181 HCP 7T and 44 NYU 3T observers, we found that most subjects had fovea-centered CMF profiles, but with significant individual differences in asymmetry and spatial concentration. K-means clustering identified five distinct CMF behavioral patterns—upper-left, lower-right, centralized, lower-centralized, and dual-centralized—which were consistently observed across both datasets.

In addition, we found a strong negative correlation between CMF and pRF size, in accordance with previous literature. The spatial distribution of $1/\sigma$ pRF estimates mirrored the CMF profiles, further validating our approach.

In summary, this framework offers a powerful tool for precise, scalable, and biologically faithful retinotopic analysis. Its application to clinical-grade fMRI data has the potential to advance visual neuroscience and inform diagnostic and rehabilitative strategies for vision disorders that affect billions worldwide.

\section{Data and Code Availability}

The data used in this research are publicly available and can be found in~\cite{benson2018hcp}. The code for processing the data and the experiment pipeline will be available upon request or \href{https://github.com/Geometric-Retinotopy-Research/Planar-CMF}{Github/Planar-CMF}.

\section{Acknowledgment}

This work was partially supported by the National Science Foundation, United States (DMS-1413417 and DMS-1412722) and the National Eye Institute, United States (R01EY032125).

\section{Declaration of Interests}
YX, ZL, and YW have a joint patent application “Xiong, Y., Z.-L. Lu, and Y. Wang, Systems and Methods for Planar Cortical Magnification Measurement”, under New York University and Arizona State University (US patent application number PCT/US2024/014617).

\section{Ethics Statement}
This study was conducted using only publicly available and previously published datasets. All data were originally collected in accordance with the ethical guidelines and institutional review board (IRB) approvals of the respective research institutions responsible for the datasets. No new data collection or participant recruitment was performed by the authors. The datasets used in this study, including those from the Human Connectome Project (HCP) and the NYU retinotopy dataset, were accessed and analyzed in full compliance with their data use agreements and terms of access.

\bibliographystyle{model2-names}
\bibliography{media_ref}

\end{document}